 \documentclass[twocolumn,10pt]{jmlr} 

\usepackage{booktabs}
\usepackage{siunitx}

\usepackage{newfloat}
\DeclareFloatingEnvironment[name={Supplementary Table},fileext=lst,listname={List of 
Supplementary Tables}]{supptable}

\usepackage{xcolor}

\theorembodyfont{\upshape}
\theoremheaderfont{\scshape}
\theorempostheader{:}
\theoremsep{\newline}

 \title[Maximizing Predictive Performance for Small Subgroups: FAIR Regression]{Maximizing Predictive Performance for Small Subgroups: Functionally Adaptive Interaction Regularization (FAIR)}

\author{
\Name{Daniel Smolyak} \Email{dsmolyak@umd.edu}\\
\addr University of Maryland, College Park, United States
\AND
\Name{Courtney Paulson} \Email{courtney.paulson@unh.edu}\\
\addr University of New Hampshire, United States
\AND
\Name{Margr\'et V. Bjarnad\'ottir} \Email{mbjarnad@umd.edu }\\
\addr University of Maryland, College Park, United States
}

\begin{document}

\maketitle

\begin{abstract}
In many healthcare settings, it is both critical to consider fairness when building analytical applications but also uniquely unacceptable to lower model performance for one group to match that of another (e.g. fairness cannot be achieved by lowering the diagnostic ability of a model for one group to match that of another and lose overall diagnostic power). Therefore a modeler needs to maximize model performance across groups as much as possible, often while maintaining a model's interpretability, which is a challenge for a number of reasons. 
In this paper we therefore suggest a new modeling framework, FAIR, to maximize performance across imbalanced groups, based on existing linear regression approaches already commonly used in healthcare settings. We propose a full linear interaction model between groups and all other covariates, paired with a weighting of samples by group size and independent regularization penalties for each group. This efficient approach overcomes many of the limitations in current approaches and manages to balance learning from other groups with tailoring prediction to the small focal group(s). FAIR has an added advantage in that it still allows for model interpretability in research and clinical settings. We demonstrate its usefulness with numerical and health data experiments.
\end{abstract}

\paragraph*{Data and Code Availability}

This paper uses the ``Diabetes 130-US hospitals for years 1999-2008'' dataset \citep{strack2014impact}, publicly available in the UCI Machine Learning Repository
. Our code is available by request.


\paragraph*{Institutional Review Board (IRB)} This study did not require IRB approval.

\section{Introduction}\label{introduction}
Healthcare providers are increasingly applying machine learning models to solve domain problems ranging from resource allocation to diagnosis and treatment decisions. Unlike traditional modeling, healthcare modeling requires a particular emphasis on fairness among outcomes for patient groups\footnote{In this paper, we will use ``group'' to refer to any distinct group in a data set for generalizability, with ``subgroup'' reserved for any further distinguished subset of an already-specified data group.}. Models must perform equally well regardless of the size or characteristics of the groups being modeled. In some domains, this issue can be resolved by pooling the entire population into one dataset, which creates a ``one size fits all'' model where any individual effects are averaged to a single effect applied to all observations. In clinical settings, pooling can have high stakes consequences, particularly when a given patient group demonstrates markedly different patient outcomes than other groups (and thus different model parameters). In this case, using a pooled model means at least one group (if not all groups) is not receiving the best possible care, which is not an acceptable approach for most healthcare settings. Instead, modelers are tasked with maximizing their model performance across each group individually to ensure best possible outcomes for all patients, a task that is often complicated by requiring models that are both simple and interpretable. 

This emphasis on straightforward usage and interpretation has led many healthcare modelers to adopt simple linear models rather than more advanced machine learning methods, even when complex models might improve overall accuracy \citep{vimont2021}. Unfortunately, maximizing performance by groups is a challenge for many reasons, even when not restricted to a linear modeling framework. First, the impact of different covariates on outcomes is often heterogeneous across groups or subgroups. For example, healthcare utilization patterns differ across demographic groups \citep{escarce1997black}, and signs of heart attack differ across genders \citep{devon2008symptoms}. Another challenge is scale. In the healthcare setting it is not unusual to have numerous possible covariates. As an example, there exist tens of thousands of individual codes for diagnoses, treatments, and drugs within health coding systems. Even after grouping codes together, prediction models often have hundreds or thousands of covariates. Finally, groups may be represented in different proportions in a dataset. Particularly in health research, there is a history of under-inclusion of marginalized racial groups in data collection \citep{nazha2019enrollment}.

These challenges complicate a critical task for healthcare modelers: how to maximize prediction accuracy across groups to account for group-specific differences without sacrificing interpretability? Several approaches have been suggested. The simplest and most common approach is to simply add group indicators to the covariates in a model \citep{goff20142013, levey2009new}. However, this falls short in linear models with differing associations between covariates and outcomes by groups (for example,  mortality outcomes for pneumonia patients according to age group). Another possibility is to fit fully separate models for each group, but this in turn falls short when there is limited data for any smaller groups.

One option to allow for flexibility in modeling these groups is to estimate all possible interactions between group identity and each covariate in the dataset. In this case, the resulting increase in the set of model covariates makes utilizing regularization important to maintain interpretability and more importantly generalizability, as well as to avoid overfitting. However, applying only standard regularization can prove insufficient due to the under-weighting of certain groups, as some covariates that may be meaningful for a small(er) group will likely be driven to zero. 

Motivated by the limitations of the above approaches, in this research we suggest a new modeling framework to increase performance across imbalanced subgroups by combining several modeling features with which healthcare researchers are already familiar. We propose the \emph{Functionally Adaptive Interaction Regularization (FAIR) approach}: a full interaction regression model between groups and all other covariates, paired with weighting of samples by group size and independent regularization penalties for each group. This approach overcomes a number of limitations of the approaches mentioned above. Most importantly, by both weighting the groups by size and independently regularizing their respective covariates, we balance learning from larger group(s) with tailored predictive accuracy on all groups. By utilizing a linear regression framework, we provide researchers with an efficient model that is familiar, easy to implement, and interpretable as well as more accurate in estimating smaller group outcomes; 
ensuring fairness in modeling across known patient groups in a setting where interpretation is paramount.

We demonstrate the relevance of our approach by comparing it to other suggested strategies available to modelers, as well as recent advances in joint regularized modeling. This includes i) fitting a separate model for each group, ii) introducing a group indicator dummy variable, and iii) a recent joint Lasso model developed by \cite{dondelinger2020joint}. We describe these models in detail in Section \ref{methods}, and we report the results on both simulated and real world data in Sections \ref{results} and \ref{realworld}, respectively. Lastly, we end with a discussion of the implications for healthcare modelers and practitioners. But first we briefly review the related literature on fairness in machine learning and regression modeling in healthcare as it relates to our study.
\section{Background}\label{background}


\subsection{Fair Machine Learning}

``Fairness'' is a relatively new concern in machine learning modeling. The rapid adoption of machine learning algorithms has been largely fueled by the dramatic improvements in prediction accuracy these computationally-intensive approaches demonstrate \citep{pessach22}. An unintended downside of improved accuracy, however, is a model that considers any improvement valuable. Thus, if biasing a model gains even a small increase in accuracy, traditional machine learning methods may incorporate or even inflate that bias to improve accuracy and thus produce potentially ``unfair'' predictions.

For some domains with low stakes problem-solving, accuracy regardless of fairness may not materially harm models \citep{mehrabi21}. Domains such as healthcare, however, require models that are able to predict fairly across groups \citep{chen2021ethical}. Healthcare practitioners and clinicians have a responsibility to ensure that diagnostic prediction algorithms perform equally well across racial groups to avoid exacerbating health disparities, for example. As a result, there is extensive research focusing on fairness between groups and subgroups of patients in the healthcare setting \citep{feng2022fair}. One approach is to directly add constraints during model training, either to prioritize fairness (defined by the authors as unbiased decision-making) or maximize fairness subject to accuracy \citep{zafar2017fairness}. These general constraints are imposed when the modeler defines the goal of the modeling, but modeling constraints can also be imposed through penalty terms in a model's objective function as well. For example, \cite{do2021joint} adds a fairness penalty term to the original sparsity and similarity penalty terms of the generalized fused Lasso model objective \citep{fusedlasso}. One example of this fused Lasso framework directly related to our work is the joint Lasso \citep{dondelinger2020joint}, which we will discuss in greater depth in Section \ref{jointlasso} as one of our direct model comparisons.

More advanced approaches include decoupling classifiers on the basis of potential performance gains for each group \citep{ustun2019fairness} and fairness considerations in feature selection \citep{belitz2021automating}. Both methods offer potential gains in group accuracy but risk decreasing overall performance. Multicalibration is an additional approach (e.g. \citep{MC1}, \citep{MC2}), but this is primarily effective when groups represent nested data, and even then can suffer from group size performance inequities \citep{ProportionalMC}. 

While these methods are all potentially valid options for addressing concerns of fairness, they share one common disadvantage: they are more complex and less interpretable than standard regression models, which have both been the traditional hallmark of healthcare modeling as well as a currently-accepted best practice \citep{wallisch2022}. We therefore next provide an overview of regression modeling in healthcare to motivate our approach in Section \ref{methods}.

\subsection{Regression Modeling in Healthcare}
In many healthcare settings, understanding why a model is making its predictions is a primary concern. Researchers building treatment or resource allocation models may ideally like high predictive accuracy, but more importantly they require model interpretability. This trade-off has resulted in traditional regression models still being used heavily in the healthcare domain despite advances in complex predictive modeling \citep{vimont2021}, especially as recent healthcare studies have shown that even when advanced models like neural networks are employed, they often demonstrate only marginal improvements over traditional regression models \citep{issitt2022}.

As a result, penalized regression models such as lasso or ridge regression are popular options. \cite{kan2019} notes the added flexibility of a regression penalty allows for better prediction of healthcare outcomes without sacrificing interpretability. Although healthcare modelers are expanding their use of penalized regression significantly, penalized regression does not directly address the issue of group-specific relationships or fairness in patient outcomes. \cite{zink2020fair} develop a health care spending regression model that penalizes separation in rates of under-compensation between subgroups. Another recent model, the joint Lasso proposed by \cite{dondelinger2020joint}, allows researchers to model identified groups separately. The formulation of the joint Lasso adds a second penalty to the differences between group-specific coefficients. This necessarily complicates interpretation and can over-penalize group disparities. We use the joint Lasso as a direct comparison method in the following sections to demonstrate both the advantages and disadvantages of their joint estimation.

\section{Methods}\label{methods}
Standard regression models are still widely used in healthcare settings due to interpretability, transparency, and simplicity. Therefore, we estimate a linear regression model in a setting with many groups, each with possibly differing relationships between covariates and the target output. We have $n$ samples with an outcome variable $y$ and feature matrix $X$ of size $n\times m$. There are $K$ non-overlapping groups ($1 \ldots K$), each of size $n_k$ samples, where $y_k$ and $X_k$ represent the outcomes and features respectively for samples in group $k$. In this section, we outline our proposed FAIR approach as an alternative to current methods, both potentially naive ``baseline'' models (representing methods readily used in practice) as well as the more advanced joint Lasso model.  

\subsection{Baseline Models}
When maximizing model performance across groups without losing the simplicity of linear regression, practitioners are often forced to rely on somewhat naive approaches based on standard regression analyses. For example, our first baseline for comparison is a ``separate models'' baseline, appropriate when groups are sufficiently distinct. Here, a penalized regression model is estimated separately for each group. In contrast, when group differences are uniform, these differences can often be modeled through a categorical variable identifying group membership. This model forms our second baseline.

\subsubsection{The joint Lasso}\label{jointlasso}

To address the pitfalls of these possibly naive methods, \cite{dondelinger2020joint} recently proposed a more sophisticated method to estimate groups jointly without sacrificing the advantages of a linear model: the joint Lasso. This method has the advantage of estimating separate coefficients for each group while penalizing differences in regression coefficients between groups to encourage group similarity.

Equation \ref{eq:joint-lasso} shows the joint Lasso objective. The $\gamma$ hyperparameter sets an overall similarity penalty, while $\tau$ allows for more precise tuning when relationships between individual groups are known in advance. Another key feature is a $\frac{1}{n_k}$ coefficient on the loss function, which ensures the loss function's ``importance'' is equal across groups of differing sizes\footnote{As in all models in this section, the $\|\|_r$ term indicates that either a lasso ($r=1$) or ridge penalty ($r=2$) may be used.}.

\begin{multline}\label{eq:joint-lasso}
\hat{\beta} = \underset{\beta=[\beta_1\ldots\beta_K]}{\arg\min} \sum_{k=1}^K \biggl\{ \frac{1}{n_k} \|y_k-X_k\beta_k\|_2^2 + \lambda\|\beta_k\|_r\\ + \gamma\sum_{k'>k} \tau_{k,k'} \|\beta_k - \beta_{k'}\|_2^2 \biggr\}
\end{multline}


\subsection{FAIR: Functionally Adaptive Interaction Regularization}

We start our proposed approach with two fundamental requirements in mind: (1) It must be extend approaches already easily applied in healthcare research, and (2) It must provide explanatory power without losing predictive power. It is worth emphasizing here that our aim is not perfect accuracy, though of course this would be ideal. 
Instead, our goal is to provide researchers with a method that provides improvements in accuracy for small groups without introducing added complexity that may be difficult to understand or interpret. For this reason, our approach adapts an interaction model approach, estimating interactions between group membership and each covariate in the data set. 

Interaction terms in regression modeling are easily understood and already implemented in many healthcare settings (e.g. drug interaction terms). We first choose an arbitrary ``large'' group as a base group. We then add interaction terms between all covariates and all group memberships (with one interaction term per group per covariate), in addition to the original covariates and group-specific intercepts. This results in a total of $m$ base coefficients, $K$ intercepts, and $m*(K-1)$ interaction coefficients. To ensure that small group size does not affect small group predictive performance, we 
incorporate the $1/n_k$ weight term from the joint Lasso. Critically, we 
allow our sparsity penalty, $\lambda_k$, to differ between groups 
\footnote{We note here that while the joint Lasso does not use interactions or shared common variables, it does take a symmetric approach by assigning each group its own coefficients.}, 
separately estimating $\beta$s for each group
. Additionally, in the joint Lasso approach, $\gamma$ is not scaled by group size and therefore does not prioritize predictive performance on smaller subgroups, which is the goal of the FAIR model. The FAIR objective function is shown in Equation \ref{eq:our-approach}. For clarification, $\beta_{k}$ indicates the coefficients for the interaction variables for group $k$, and $\beta_1$ indicates the base group coefficients.

\begin{multline}\label{eq:our-approach}
\hat{\beta} = \underset{\beta=[\beta_1\ldots\beta_K]}{\arg\min} \sum_{k=2}^K \left\{\frac{1}{n_k} \|y_k-X_k(\beta_1+\beta_k)\|_2^2 \right\} + \\ \frac{1}{n_1} \|y_1-X_1\beta_1\|_2^2 + \sum_{k=1}^K \lambda_k\|\beta_k\|_r
\end{multline}


While our methodology is agnostic to choice of sparsity penalty, for consistency and simplicity of presentation, we focus on the L1 (or Lasso) penalty for this model and all other comparison models.



\subsection{Evaluating Model Performance}

Because this paper focuses on predicting numerical outcomes, we evaluate all models' performance using the Mean Squared Error (MSE). A lower score indicates increased prediction accuracy. We focus particularly on performance within our designated ``small'' group, given our previously stated goals of accuracy and fairness. Performance for the large group is less difficult to maximize – running a model on the entire dataset is often sufficient, and the performance for the large group can be independently maximized across different approaches.

Further, because all methods used for comparison here are estimated using an L1 regularization, we ensure fair evaluation by conducting a cross-validated grid search within the training data over a wide range of potential values for all tuning parameters for each method. The combination that results in the best performance on average for the small group(s) across the folds for each method is chosen for use on a (hold-out) testing dataset.

\section{Numerical Experiments}\label{results}
We begin our evaluation of model performance by simulating numerical experiments. For simplicity, 
we primarily focus on two independently-generated groups, one large and one small. We utilize simulated data 
to demonstrate FAIR's performance for small group prediction under several different data conditions. We compare this performance to the baselines and joint Lasso models discussed above to provide prescriptive suggestions for situations in which healthcare researchers might see advantages by modeling with the FAIR approach. 
For further details on the underlying data generation used here, please refer to Section \ref{parameters} in the Appendix.

\subsection{Results: The Base Case}\label{sec:ours-excels}
We first define a base case as follows: the large group size is set to $n_1=300$, the small group size to $n_2=100$, and the number of features to $m=30$. Of those features, 20 are set to zero (in other words, uninformative) for both groups. For the large group the remaining ten coefficients are set to one, while for the small group, seven of the remaining coefficients are set to one and three coefficients are set to three (thus three coefficients differ between the groups). Both groups are generated with output noise ($\epsilon_k$, specified in equation \ref{eq:sim} in Appendix \ref{parameters}) equal to one.

\begin{figure}[!htbp]
    \centering
    \includegraphics[width=7cm]{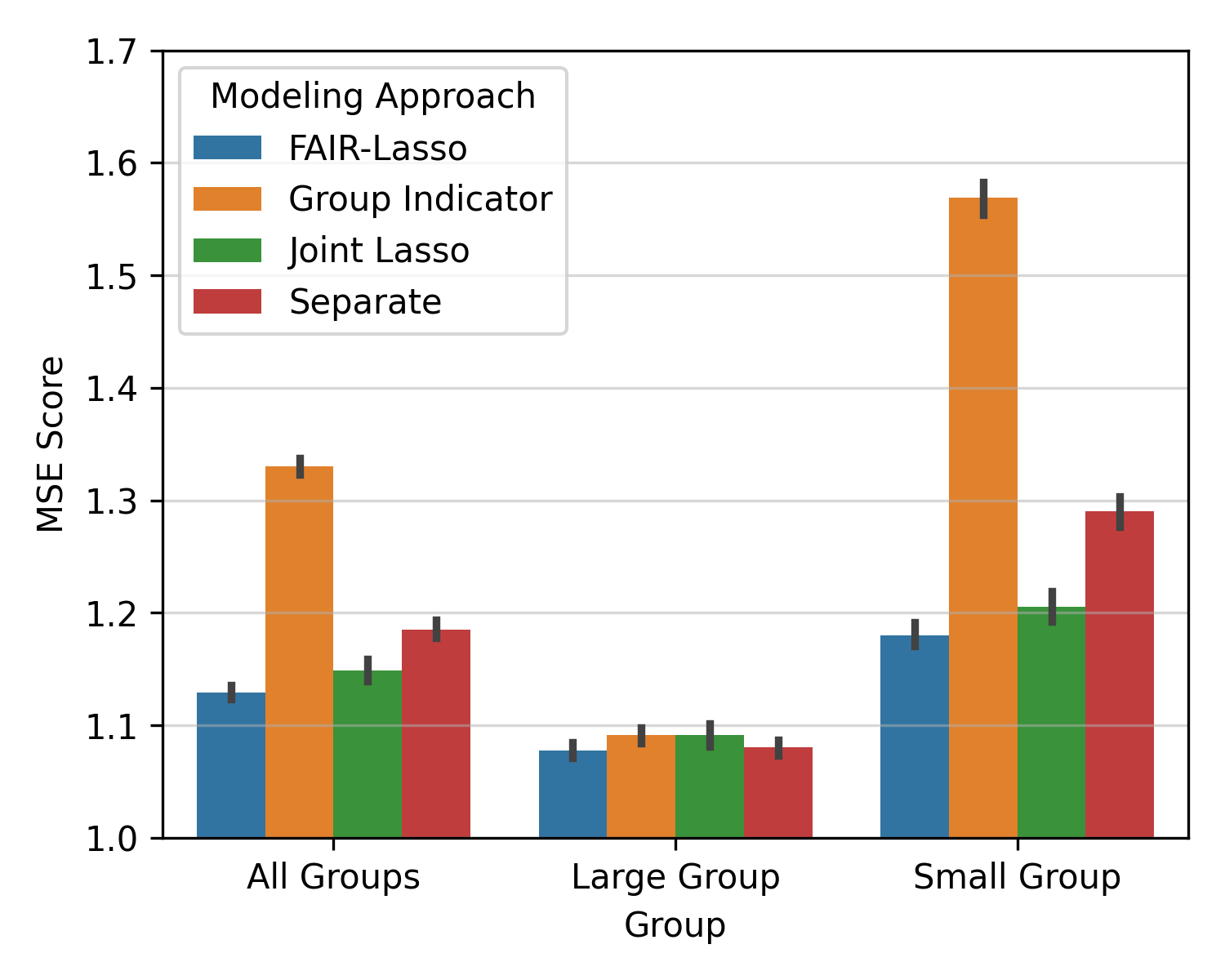}
    \caption{Performance of all modeling approaches on the base case, using a testing set of 1000 samples per group, including 95\% confidence interval across 250 iterations.}
    \label{fig:best-perf-sim}
\end{figure}

Figure \ref{fig:best-perf-sim} summarizes the results of the FAIR approach, the baseline models, and the joint Lasso. 
The base case represents the motivating premise for the development of FAIR: data with groups that both share information but also demonstrate clear deviation. Due to these significant differences, the group indicator baseline performs poorly when trying to capture differences in the small group through only categorical group indicator variables; the separate models baseline performs well on the large group but suffers on the small group with less data. 
By using interactions and separate penalty parameters for each group, FAIR highlights the benefit of its added flexibility in modeling. The joint Lasso performs similarly (but slightly worse) compared to our method, as it also enables independent coefficient setting and information sharing between the two groups.

\subsection{Results: Varying Simulation Parameters}\label{sec:sim-varied}
\begin{figure*}[!htbp]
    \centering
    \includegraphics[scale=0.45]{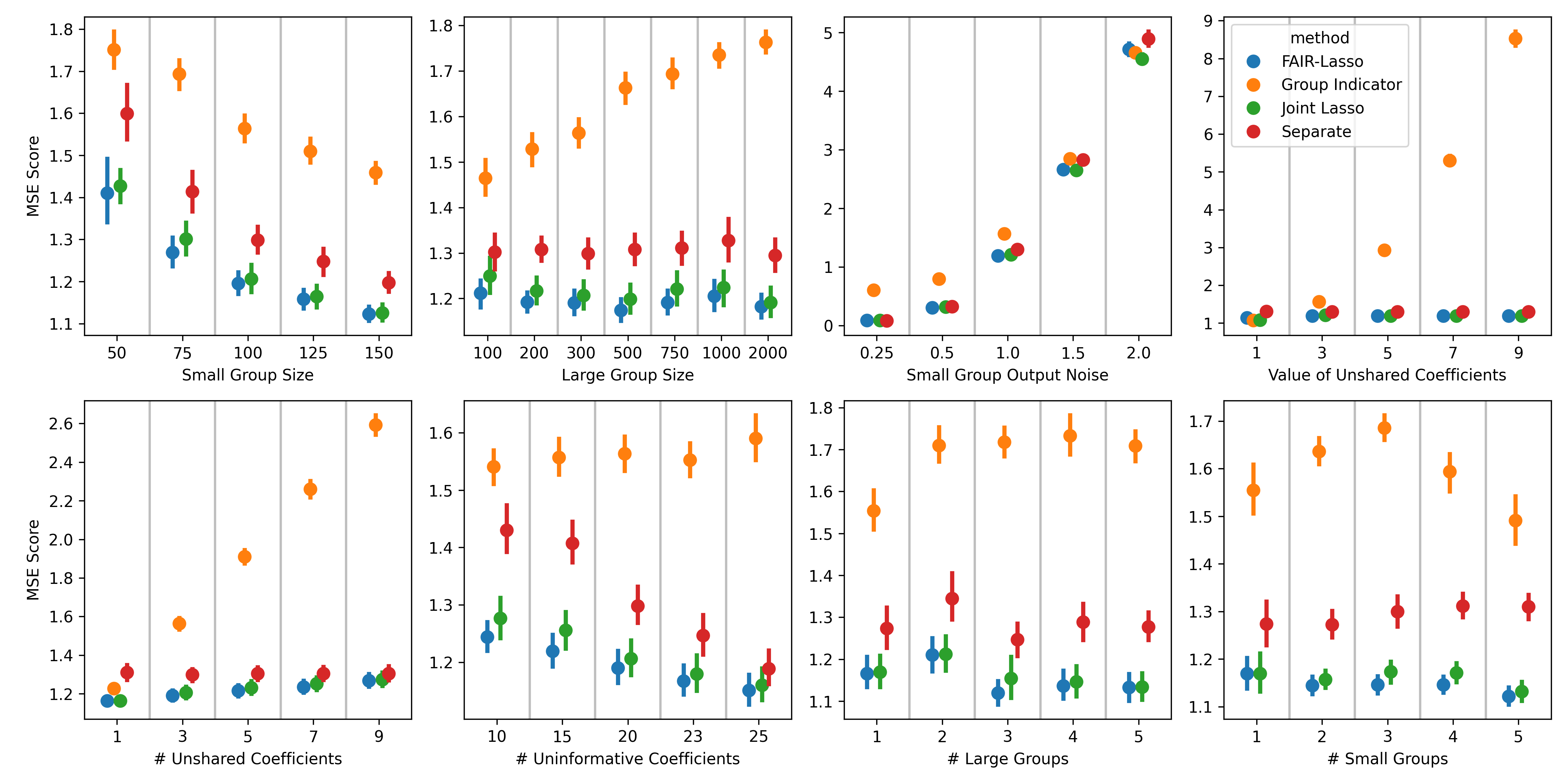}
    \caption{Performance of the FAIR approach, the baseline models and joint Lasso on the small group as each parameter (indicated by the x-axis label) is varied - all other parameters held constant.}
    \label{fig:lever-linear}
\end{figure*}

Although the base case demonstrates the motivation behind FAIR, we recognize such an exact trade-off in parameters may be unlikely in practice. To that end, we expand around that base case by varying the parameters used to simulate the data and provide a more holistic look at FAIR's performance relative to the comparison models. The results of varying the parameters to their relevant extremities are summarized in Figure \ref{fig:lever-linear}. For more details on these parameters, please refer to Appendix \ref{parameters} and Table \ref{tab:sim_params}.

Overall, we see two clear patterns: (1) FAIR consistently outperforms the two baseline methods, and (2) FAIR also outperforms the joint Lasso, but often within the margin of error. Beginning from the top-left plot, we see that as small group size increases, performance improves for all methods, as expected. However, FAIR performs the best even at larger group sizes, with joint Lasso a close second.
Next, as large group size changes, FAIR's MSE stays consistently low, generally significantly lower than the other methods (except for joint Lasso).
The group indicator baseline performs worse as the large group grows, given the relative decreasing size of the small group. 
Examining the small group output noise, we note that FAIR, the joint Lasso, and separate models all perform well when the noise is small; small noise makes it easier to learn the ``truth'' from just 100 observations. However, this also means as small group noise increases, MSE begins to increase correspondingly in all methods due to the small sample size. 
Moving to the fourth parameter in our simulated experiments, differences in the value of coefficients that are not shared between the two groups (top right), we see when the coefficients for both groups are equal to 1, the comparison models perform similarly to FAIR, but as the magnitude of the differences increases, the group indicator lags by failing to account for the impact of different covariates on different groups. These results parallel the results of varying the number of unshared coefficients (bottom left). As the number of unshared coefficients and thus group differences increases, FAIR's performance converges with separate models (as there is less to learn from the large group), while the MSE of 
the group indicator diverges. 

Next we vary the number of uninformative coefficients. We see the performance of separate models converges towards FAIR's performance as the number of uninformative coefficients increases. As the number of uninformative coefficients increases, more covariate relationships are set to zero; while this is technically shared information between groups, it is not useful given it indicates no relationship. 
Lastly, we vary the total number of groups, first varying the number of ``large'' groups (those of size 300) and then varying the number of ``small'' groups (those of size 100). FAIR continues to consistently perform best. 

In summary, regardless of parameters, FAIR continues to perform well. It is at least comparable if not demonstrably better than each of the comparison methods across all eight dimensions tested. Interestingly, joint Lasso performs on par with FAIR in just about every setting, demonstrating the relative similarity of the two methods. However, as discussed in the following section, FAIR does have two advantages over the joint Lasso: speed and implementation. 
This indicates FAIR as is a good choice for general modeling when the characteristics of data are both known (e.g. the size of each subgroup) and unknown (e.g. the noise present in the dataset). Even with large deviations from the expected use case, FAIR's relative prediction accuracy does not suffer.

\subsection{Speed and Implementation Advantages}

In addition to the improved performance of FAIR compared to joint Lasso, FAIR also yields advantages both in ease of implementation and algorithm speed. The FAIR method fits within the implementation of the popular glmnet R package \citep{hastie2021introduction}, as the glmnet function allows for specification of individual sample weights and sparsity penalties (the \texttt{weights} and \texttt{penalty\_factor} parameters). This allows healthcare users to adapt a package and method already in use to improve small group accuracy.

Perhaps more compellingly, this implementation of FAIR is 19 times faster than joint Lasso in our base case simulations, even using the provided joint Lasso R \texttt{fuser} package. For more on the timing trials for both the base case simulations and the following real data example, please see Appendix \ref{apd:timing}.


\section{Real World Experiment}\label{realworld}

\begin{figure}[!htbp]
    \centering
    \includegraphics[width=7cm]{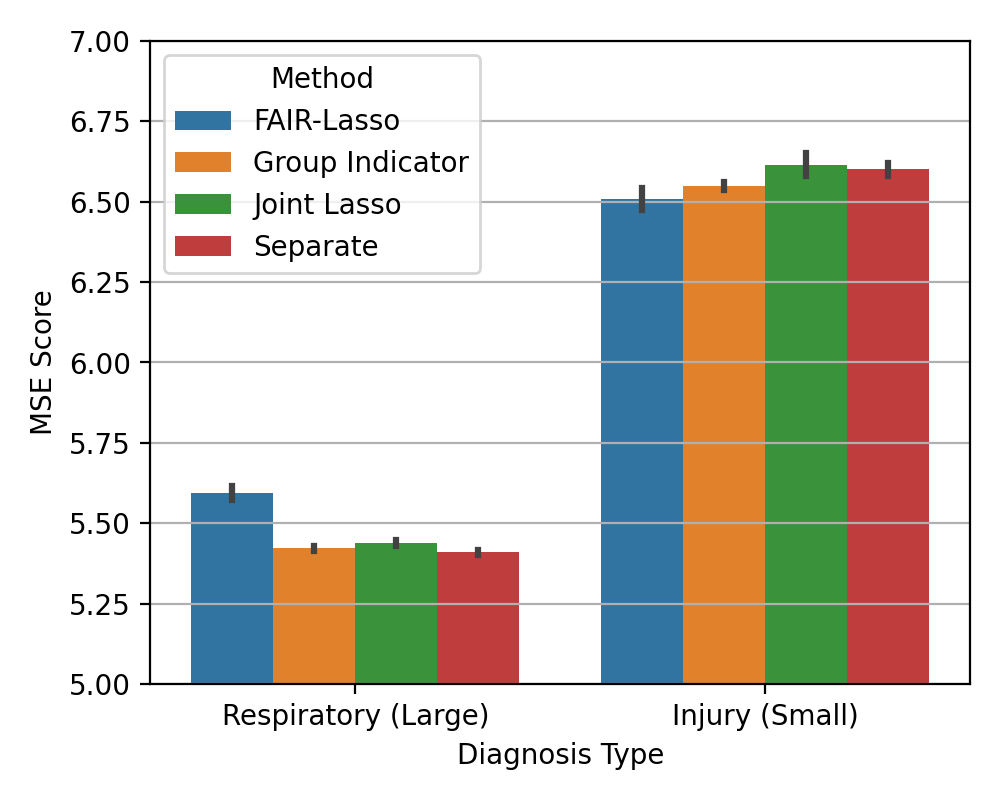}
    \caption{Model testing performance on length of stay prediction, including 95\% confidence interval on 250 random samples.}
    \label{fig:best-perf-real}
\end{figure}

To demonstrate a use case of both FAIR and the comparison models in practice, we apply the models to the ``Diabetes 130-US hospitals for years 1999-2008'' dataset \citep{strack2014impact}. This dataset contains extensive information on diabetic patients' inpatient stays, including length of stay, admission source, 
number of procedures, and primary diagnosis.
Of note, while the data collected is on diabetes patients specifically, diabetes may not be the main reason for hospitalization (i.e., their primary diagnosis). For this use case, we model length of stay and use primary diagnosis to differentiate our groups. This reflects the likelihood that length of stay effects will differ across covariates 
depending on the primary diagnosis. For instance, the number of procedures may predict length of stay differently for an injury diagnosis than for a respiratory diagnosis, given the varying levels of complexity in the involved procedures. However, many variables likely have similar effects even across diagnoses, such as patient age. 
Thus, we expect our method to facilitate information sharing for some variables while allowing other coefficients to diverge where appropriate.

To simulate the data scarcity common in many healthcare settings, we take random samples of the full dataset. Specifically, we create one ``large'' group, patients with a ``Respiratory'' primary diagnosis, and one ``small'' group, patients with an ``Injury'' primary diagnosis. We sample 2000 patients for the large group and 200 patients for the small group to build a training dataset. For robustness, 2000 patients are sampled from each group for the testing dataset.

As shown in Figure \ref{fig:best-perf-real}, here FAIR outperforms both the  baseline models and the joint LASSO for the small group (injury patients). While FAIR does not perform best for the large group, as mentioned above, FAIR's emphasis is on predicting the small group, while the large group prediction can be independently optimized (e.g. with its own model). 

\begin{figure*}[!htbp]
    \centering
    \includegraphics[scale=.5]{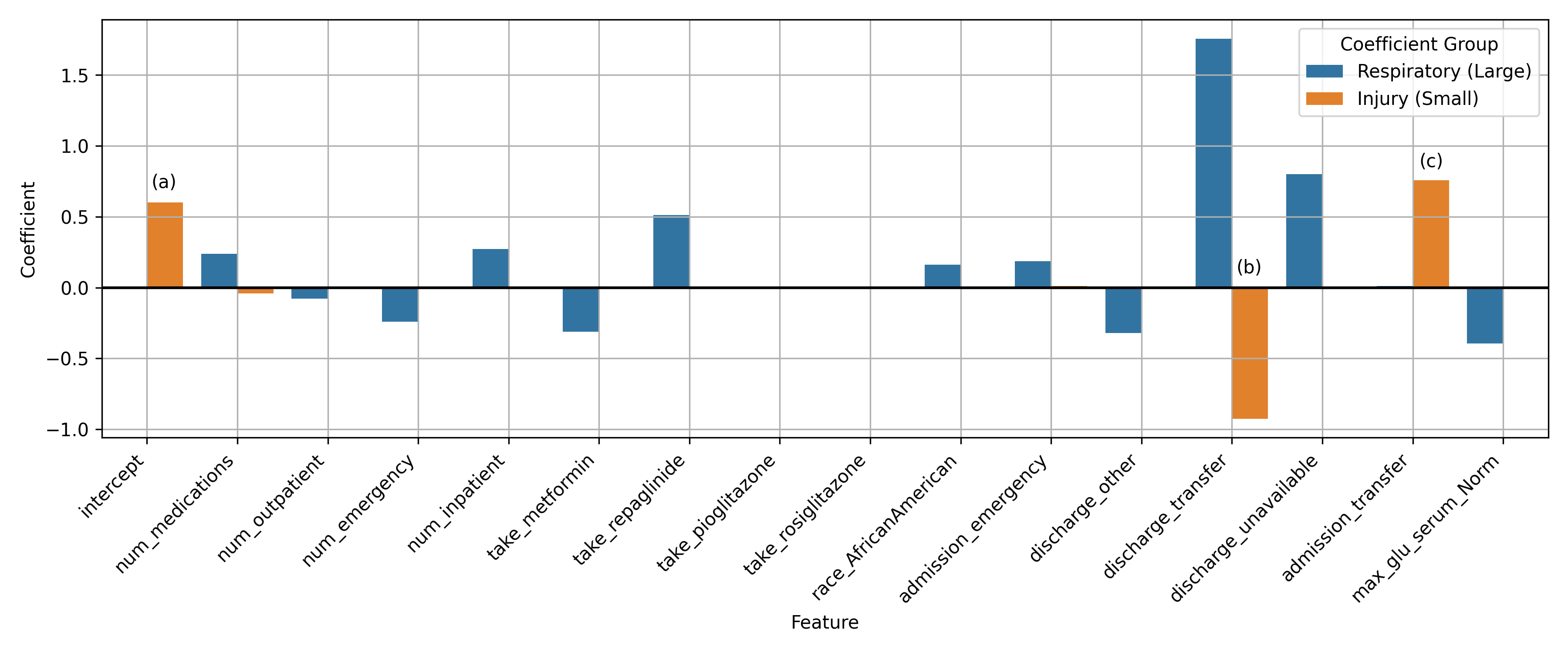}
    \caption{Coefficients of FAIR-Lasso on the Diabetes Hospital Dataset. Blue bars indicate the coefficients of the base/large group (``Respiratory'' patients) and orange for the coefficients of the small group (``Injury'' patients).}
    \label{fig:uci-coefficients}
\end{figure*}

While model performance is important, a major factor in our choice of an adapted regression model is the interpretability of the model. In healthcare settings, the explanatory power of a model is often of equal or greater importance to the predictive power. Figure \ref{fig:uci-coefficients} shows a summary of the coefficients resulting from the FAIR-Lasso model in this exemplar setting.
(Please refer to Appendix \ref{apd:first} for a complete description of the variables shown in Figure \ref{fig:uci-coefficients}.)

Many of the interaction coefficients for the small group are zero, but for several features, the small group either adds a new effect, such as the Injury-specific intercept or an admission by transfer (shown in Figure \ref{fig:uci-coefficients} as (``a'') and (``c'')) or mitigates the effect of the base coefficient, such as patients discharged by transfer (``b''). Each of these is an effect with a clear clinical understanding. For (a), we expect an increase in the intercept for the small group, because Injury patients have a higher average length of stay in the data compared to Respiratory patients (4.5 days vs. 4 days). For (b), a discharge to transfer may indicate poor health for both groups, but Respiratory patients require longer and more careful monitoring before a transfer decision.
Finally, for (c), an admission from transfer may indicate a more serious injury that requires increased care and recovery in the hospital, while for Respiratory patients, there may be no difference in severity for patients admitted by transfer. 

\section{Discussion}\label{discussion}
In this paper we introduce a novel modeling approach, FAIR, that emphasizes maximizing predictive performance on smaller groups while maintaining 
interpretability through an extended linear interaction model. We argue this approach may be helpful in numerous scenarios in healthcare, especially considering the often small groups present in healthcare data. We note the FAIR approach can be applied across important grouping covariates like demographics. FAIR performs particularly well when the number of different and common effects by groups is balanced, as well as when small groups in the data are considerably smaller and thus difficult to accurately estimate based solely on their own data.

Our approach does have some limitations, however. Where groups are either very dissimilar or very similar, the FAIR approach is unnecessary. Separate models will capture very dissimilar effects, while using simple group indicator variables adjusts for minor differences when group covariate effects are similar. Likewise, when healthcare professionals have sufficiently large data for all groups, FAIR may not be necessary. Further, when modelers wish only to achieve the highest possible prediction accuracy, FAIR would not be advised; its emphasis on interpretability may not outweigh the accuracy gains possible from advanced machine learning models.

Our extensive numerical experiments attempt to capture the most relevant dimensions in which data could vary. However, future work includes exploring other parameters, e.g. correlations between covariates. An interesting future direction is 
to apply the methods at scale along multiple dimensions as well, including differences when easily translating FAIR to binary classification using regularized logistic regression. 
Further, when considering baseline models for comparison, we chose to compare FAIR to other regression models to highlight interpretable approaches, but as a consequence we have left out other relevant approaches. In addition to previously discussed approaches such as multilevel modeling and multicalibration, other methods such as transfer learning also share information from better represented groups with less represented groups \citep{li2021targeting, gu2022transfer, bastani2021predicting}. Additionally, while we used 
MSE as a performance metric, other metrics are important 
depending on the modeling context. For example, a hospital length of stay model may be used to rank patients for their need of case management. In that context, measuring how each approach impacts allocation of case management resources across groups is also critical.

\bibliography{bibliography}

\begin{thebibliography}{28}
\providecommand{\natexlab}[1]{#1}
\providecommand{\url}[1]{\texttt{#1}}
\expandafter\ifx\csname urlstyle\endcsname\relax
  \providecommand{\doi}[1]{doi: #1}\else
  \providecommand{\doi}{doi: \begingroup \urlstyle{rm}\Url}\fi

\bibitem[Bastani(2021)]{bastani2021predicting}
Hamsa Bastani.
\newblock Predicting with proxies: Transfer learning in high dimension.
\newblock \emph{Management Science}, 67\penalty0 (5):\penalty0 2964--2984,
  2021.

\bibitem[Belitz et~al.(2021)Belitz, Jiang, and Bosch]{belitz2021automating}
Clara Belitz, Lan Jiang, and Nigel Bosch.
\newblock Automating procedurally fair feature selection in machine learning.
\newblock In \emph{Proceedings of the 2021 AAAI/ACM Conference on AI, Ethics,
  and Society}, pages 379--389, 2021.

\bibitem[Chen et~al.(2021)Chen, Pierson, Rose, Joshi, Ferryman, and
  Ghassemi]{chen2021ethical}
Irene~Y Chen, Emma Pierson, Sherri Rose, Shalmali Joshi, Kadija Ferryman, and
  Marzyeh Ghassemi.
\newblock Ethical machine learning in healthcare.
\newblock \emph{Annual review of biomedical data science}, 4:\penalty0
  123--144, 2021.

\bibitem[DeVon et~al.(2008)DeVon, Ryan, Ochs, and Shapiro]{devon2008symptoms}
Holli~A DeVon, Catherine~J Ryan, Amy~L Ochs, and Moshe Shapiro.
\newblock Symptoms across the continuum of acute coronary syndromes:
  differences between women and men.
\newblock \emph{American journal of critical care}, 17\penalty0 (1):\penalty0
  14--24, 2008.

\bibitem[Do et~al.(2021)Do, Nandi, Putzel, Smyth, and Zhong]{do2021joint}
Hyungrok Do, Shinjini Nandi, Preston Putzel, Padhraic Smyth, and Judy Zhong.
\newblock Joint fairness model with applications to risk predictions for
  under-represented populations.
\newblock \emph{arXiv preprint arXiv:2105.04648}, 2021.

\bibitem[Dondelinger et~al.(2020)Dondelinger, Mukherjee, and
  Initiative]{dondelinger2020joint}
Frank Dondelinger, Sach Mukherjee, and Alzheimer’s Disease~Neuroimaging
  Initiative.
\newblock The joint lasso: high-dimensional regression for group structured
  data.
\newblock \emph{Biostatistics}, 21\penalty0 (2):\penalty0 219--235, 2020.

\bibitem[Escarce and Puffer(1997)]{escarce1997black}
Jos{\'e}~J Escarce and Frank~W Puffer.
\newblock \emph{Black-white differences in the use of medical care by the
  elderly: a contemporary analysis}.
\newblock Washington, DC: National Academy Press, 1997.

\bibitem[Feng et~al.(2022)Feng, Du, Zou, and Hu]{feng2022fair}
Qizhang Feng, Mengnan Du, Na~Zou, and Xia Hu.
\newblock Fair machine learning in healthcare: A review.
\newblock \emph{arXiv preprint arXiv:2206.14397}, 2022.

\bibitem[Globus-Harris et~al.(2023)Globus-Harris, Harrison, Kearns, Roth, and
  Sorrell]{MC2}
Ira Globus-Harris, Declan Harrison, Michael Kearns, Aaron Roth, and Jessica
  Sorrell.
\newblock Multicalibration as boosting for regression.
\newblock In \emph{Proceedings of the 40th International Conference on Machine
  Learning}, pages 11459–--11492, 2023.

\bibitem[Goff~Jr et~al.(2014)Goff~Jr, Lloyd-Jones, Bennett, Coady,
  D’agostino, Gibbons, Greenland, Lackland, Levy, O’donnell,
  et~al.]{goff20142013}
David~C Goff~Jr, Donald~M Lloyd-Jones, Glen Bennett, Sean Coady, Ralph~B
  D’agostino, Raymond Gibbons, Philip Greenland, Daniel~T Lackland, Daniel
  Levy, Christopher~J O’donnell, et~al.
\newblock 2013 acc/aha guideline on the assessment of cardiovascular risk: a
  report of the american college of cardiology/american heart association task
  force on practice guidelines.
\newblock \emph{Circulation}, 129\penalty0 (25\_suppl\_2):\penalty0 S49--S73,
  2014.

\bibitem[Gu et~al.(2022)Gu, Han, and Duan]{gu2022transfer}
Tian Gu, Yi~Han, and Rui Duan.
\newblock A transfer learning approach based on random forest with application
  to breast cancer prediction in underrepresented populations.
\newblock In \emph{PACIFIC SYMPOSIUM ON BIOCOMPUTING 2023: Kohala Coast,
  Hawaii, USA, 3--7 January 2023}, pages 186--197. World Scientific, 2022.

\bibitem[Hastie et~al.(2021)Hastie, Qian, and Tay]{hastie2021introduction}
Trevor Hastie, Junyang Qian, and Kenneth Tay.
\newblock An introduction to glmnet.
\newblock \emph{CRAN R Repositary}, 5:\penalty0 1--35, 2021.

\bibitem[Hebert-Johnson et~al.(2018)Hebert-Johnson, Kim, Reingold, and
  Rothblum]{MC1}
Ursula Hebert-Johnson, Michael~P. Kim, Omer Reingold, and Guy~N. Rothblum.
\newblock Multicalibration: Calibration for the (computationally-identifiable)
  masses.
\newblock In \emph{Proceedings of the 35th International Conference on Machine
  Learning, Stockholm, Sweden, PMLR 80}, 2018.

\bibitem[Issitt et~al.(2022)Issitt, Cortina-Borja, Bryant, Bowyer, Taylor, and
  Sebire]{issitt2022}
Richard~W. Issitt, Mario Cortina-Borja, William Bryant, Stuart Bowyer,
  Andrew~M. Taylor, and Neil Sebire.
\newblock Classification performance of neural networks versus logistic
  regression models: Evidence from healthcare practice.
\newblock \emph{Cureus}, 14\penalty0 (2), 2022.

\bibitem[Kan et~al.(2019)Kan, Kharrazi, Chang, Bodycombe, Lemke, and
  Weiner]{kan2019}
Hong~J. Kan, Hadi Kharrazi, Hsien-Yen Chang, Dave Bodycombe, Klaus Lemke, and
  Jonathan~P. Weiner.
\newblock Exploring the use of machine learning for risk adjustment: A
  comparison of standard and penalized linear regression models in predicting
  health care costs in older adults.
\newblock \emph{PLoS One}, March 2019.

\bibitem[La~Cava et~al.(2023)La~Cava, Lett, and Wan]{ProportionalMC}
William~G. La~Cava, Elle Lett, and Guangya Wan.
\newblock Fair admission risk prediction with proportional multicalibration.
\newblock \emph{Proceedings of Machine Learning Research}, 209:\penalty0
  350–--378, 2023.

\bibitem[Levey et~al.(2009)Levey, Stevens, Schmid, Zhang, Castro~III, Feldman,
  Kusek, Eggers, Van~Lente, Greene, et~al.]{levey2009new}
Andrew~S Levey, Lesley~A Stevens, Christopher~H Schmid, Yaping Zhang,
  Alejandro~F Castro~III, Harold~I Feldman, John~W Kusek, Paul Eggers,
  Frederick Van~Lente, Tom Greene, et~al.
\newblock A new equation to estimate glomerular filtration rate.
\newblock \emph{Annals of internal medicine}, 150\penalty0 (9):\penalty0
  604--612, 2009.

\bibitem[Li et~al.(2021)Li, Cai, and Duan]{li2021targeting}
Sai Li, Tianxi Cai, and Rui Duan.
\newblock Targeting underrepresented populations in precision medicine: A
  federated transfer learning approach.
\newblock \emph{arXiv preprint arXiv:2108.12112}, 2021.

\bibitem[Mehrabi et~al.(2021)Mehrabi, Morstatter, Saxena, Lerman, and
  Galstyan]{mehrabi21}
Ninareh Mehrabi, Fred Morstatter, Nripsuta Saxena, Kristina Lerman, and Aram
  Galstyan.
\newblock A survey on bias and fairness in machine learning.
\newblock \emph{ACM Computing Surveys}, 54\penalty0 (6), July 2021.

\bibitem[Nazha et~al.(2019)Nazha, Mishra, Pentz, and
  Owonikoko]{nazha2019enrollment}
Bassel Nazha, Manoj Mishra, Rebecca Pentz, and Taofeek~K Owonikoko.
\newblock Enrollment of racial minorities in clinical trials: old problem
  assumes new urgency in the age of immunotherapy.
\newblock \emph{American Society of Clinical Oncology Educational Book},
  39:\penalty0 3--10, 2019.

\bibitem[Pessach and Shmueli(2022)]{pessach22}
Dana Pessach and Erez Shmueli.
\newblock A review on fairness in machine learning.
\newblock \emph{ACM Computing Surveys}, 55\penalty0 (3), Ferbruary 2022.

\bibitem[Strack et~al.(2014)Strack, DeShazo, Gennings, Olmo, Ventura, Cios,
  Clore, et~al.]{strack2014impact}
Beata Strack, Jonathan~P DeShazo, Chris Gennings, Juan~L Olmo, Sebastian
  Ventura, Krzysztof~J Cios, John~N Clore, et~al.
\newblock Impact of hba1c measurement on hospital readmission rates: analysis
  of 70,000 clinical database patient records.
\newblock \emph{BioMed research international}, 2014, 2014.

\bibitem[Tibshirani and Saunders(2005)]{fusedlasso}
Robert Tibshirani and Michael Saunders.
\newblock Sparsity and smoothness via the fused lasso.
\newblock \emph{Journal of the Royal Statistical Society, Series B},
  67:\penalty0 91--108, 2005.

\bibitem[Ustun et~al.(2019)Ustun, Liu, and Parkes]{ustun2019fairness}
Berk Ustun, Yang Liu, and David Parkes.
\newblock Fairness without harm: Decoupled classifiers with preference
  guarantees.
\newblock In \emph{International Conference on Machine Learning}, pages
  6373--6382. PMLR, 2019.

\bibitem[Vimont et~al.(2022)Vimont, Leleu, and Durand-Zaleski]{vimont2021}
Alexandre Vimont, Henri Leleu, and Isabelle Durand-Zaleski.
\newblock Machine learning versus regression modelling in predicting individual
  healthcare costs from a representative sample of the nationwide claims
  database in france.
\newblock \emph{European Journal of Health Economics}, 23:\penalty0 211--223,
  2022.

\bibitem[Wallisch et~al.(2022)Wallisch, Bach, Hafermann, Klein, Sauerbrei,
  Steyerberg, Heinze, and Rauch]{wallisch2022}
Christine Wallisch, Paul Bach, Lorena Hafermann, Nadja Klein, Willi Sauerbrei,
  Ewout~W. Steyerberg, Georg Heinze, and Geraldine Rauch.
\newblock Review of guidance papers on regression modeling in statistical
  series of medical journals.
\newblock \emph{PLoS One}, 17\penalty0 (1), January 2022.

\bibitem[Zafar et~al.(2017)Zafar, Valera, Gomez~Rodriguez, and
  Gummadi]{zafar2017fairness}
Muhammad~Bilal Zafar, Isabel Valera, Manuel Gomez~Rodriguez, and Krishna~P
  Gummadi.
\newblock Fairness beyond disparate treatment \& disparate impact: Learning
  classification without disparate mistreatment.
\newblock In \emph{Proceedings of the 26th international conference on world
  wide web}, pages 1171--1180, 2017.

\bibitem[Zink and Rose(2020)]{zink2020fair}
Anna Zink and Sherri Rose.
\newblock Fair regression for health care spending.
\newblock \emph{Biometrics}, 76\penalty0 (3):\penalty0 973--982, 2020.

\end{thebibliography}

\appendix

\section{Simulation Parameters}\label{parameters}

To generate the simulated data, for each group $k$ we sample a feature matrix $X_k$ from a standard normal distribution $\mathcal{N}(0,1)$. For each group, the outcome $y_k$ is calculated as:
\begin{equation} \label{eq:sim}
y_k = \beta_k X_k + \epsilon_k,
\end{equation}
where $\beta_k$ is a vector of coefficients, and $\epsilon_k$ is 
Gaussian noise sampled from a normal distribution ($\mathcal{N}(0,\sigma_{\epsilon}^2$); $\beta_k$ and $\epsilon_k$ vary across the simulation scenarios. We use a zero covariance matrix to generate the features, instead relating distinct groups through their relationship with the outcome by sharing values for $\beta_k$.

\begin{table*}[]
\centering
\begin{tabular}{|l|l|}
    \hline
    \textbf{Parameter} & \textbf{Description}  \\ \hline
    \# Covariates & Number of covariates \\
    Group Size & Number of samples in each group \\
    Group Output Noise & Variance of Gaussian noise ($\epsilon_k$) added in equation \ref{eq:sim} for each group  \\
    \# Uninformative Coefficients & Number of coefficients set to 0 \\
    Default Coefficient & Default value of each informative coefficient \\
    \# Unshared Coefficients & Number of coefficients differing between groups \\
    Value of Unshared Coefficient & Value of the coefficients in small group that differ from large group \\
    \# Small Groups & Number of groups with ``small'' number of samples \\
    \# Large Groups & Number of groups with ``large'' number of samples \\ \hline
\end{tabular}
\caption{Parameters in our simulation framework.}
\label{tab:sim_params}
\end{table*}

We summarize all of the data/modeling parameters in Table \ref{tab:sim_params}. Our simulation aims to demonstrate performance as we change eight of these parameters: (1) the size of the small group, (2) the size of the large group, (3) the noise present in the generation of the small group data, (4) the relative size difference in the coefficients which are not shared between the small and large groups, (5) the number of coefficients which differ between the small and large groups, the number of uninformative coefficients in the data (that is, the number of coefficients in the data that should be set to 0 by the penalized regression), (7) the number of large groups, and (8) the number of small groups.

All parameters are adjusted one at a time from a base case. We set our base case described in Section \ref{sec:ours-excels}: $n_1=300$, $n_2=100$, and $m=30$. Of those features, 20 are set to zero for both groups. For the large group the remaining ten coefficients are set to one; for the small group, seven of the remaining coefficients are also set to one and three coefficients are set to three. Both groups were generated with output noise ($\epsilon_k$) equal to one.

Section \ref{sec:ours-excels} demonstrates the base case. Section \ref{sec:sim-varied} demonstrates simulated results when each of the eight parameters above are varied, 
changing only one parameter at a time to show the direct effect of the variation on model performance.

\section{Diabetes Hospital Dataset Variables}\label{apd:first}

We preprocess the ``Diabetes 130-US hospitals for years 1999-2008'' dataset in order to prepare it for regression modeling. The resulting variables and their descriptions are in Supplementary Table \ref{tab:uci_var_descriptions}. For example, the $age$ variable is originally provided in 10-year buckets – we select the lower end of the range to convert $age$ to an integer format. For the medication data, we remove some of the rarely occurring medications, and simplify each remaining medication to a Boolean variable indicating whether or not the medication is prescribed (and do not incorporate the information about changes in dosage). To view the full set of preprocessing steps, the dataset preprocessing code is available in the aforementioned code repository, as well as the preprocessed dataset.

\begin{supptable*}[]
    \centering
    \begin{tabular}{|l|l|} \hline
        \textbf{Feature Name} & \textbf{Description} \\ \hline
        \textit{num\_lab\_procedures} & Number of lab tests performed during the encounter \\
        \textit{num\_procedures} & Number of procedures (other than lab tests) performed during the encounter \\
        \textit{num\_medications} & Number of distinct generic names administered during the encounter \\
        \textit{num\_outpatient} & Number of outpatient visits of the patient in the year preceding the encounter \\
        \textit{num\_emergency} & Number of emergency visits of the patient in the year preceding the encounter \\
        \textit{num\_inpatient} & Number of inpatient visits of the patient in the year preceding the encounter \\
        \textit{num\_diagnoses} & Number of diagnoses entered to the system \\
        \textit{change} & Indicates if there was a change in diabetic medications \\
        \textit{diabetesMed} & Indicates if there was any diabetic medications prescribed \\
        \textit{take\_[blank]}& Indicates if the $blank$ drug was prescribed \\
        \textit{race\_[blank]} & Race (Values: Caucasian, Asian, African American, Hispanic, Other) \\
        \textit{gender\_Male} & Gender, Male = 1 \\
        \textit{admission\_type} & Admission type (Values: emergency, unavailable) \\
        \textit{discharge} & Discharge disposition (Values: transfer, other, unavailable) \\
        \textit{admission\_source} & Admission source (Values: referral, transfer, unavailable) \\
        \textit{max\_glu\_serum\_Norm} & Indicates whether the result of a glucose serum test was in the ``normal'' range \\
        \textit{A1Cresult\_Norm} & Indicates whether the result of a A1C test was in the ``normal'' range \\ \hline
    \end{tabular}
    \caption{Feature names and descriptions from the Diabetes Hospital dataset.}
    \label{tab:uci_var_descriptions}
\end{supptable*}

\section{Method Speed Comparison}\label{apd:timing}

As mentioned in the main text, our method is many times faster than joint Lasso, specifically 19 times faster in our simulation base case and 10 times faster in the Hospital dataset. Total time to run was averaged across all 250 trials in both scenarios.

\begin{supptable*}
\centering
\begin{tabular}{lrr}
\toprule
Method & Hospital & Simulation \\
\midrule
FAIR Lasso & 1.189375 & 0.999165 \\
Group Indicator & 0.381009 & 0.368708 \\
Joint Lasso & 11.783740 & 19.084862 \\
Separate & 0.489743 & 0.092180 \\
\bottomrule
\end{tabular}
\caption{Time in seconds (averaged across 250 runs) for each method on each dataset.}
\label{tab:my_label}
\end{supptable*}

\end{document}